\documentclass[shortnote]{jpsj3}
\usepackage{txfonts}
\usepackage{color}
\usepackage{ulem}

\title{
Suppression of Structural Phase Transition in IrTe$_2$ by Isovalent Rh Doping
}

\author
{\name{Kazutaka \surname{Kudo}}\thanks{E-mail: kudo@science.okayama-u.ac.jp},
\name{Masakazu \surname{Kobayashi}},
\name{Sunseng \surname{Pyon}}\thanks{Present address: Department of Applied Physics, The University of Tokyo, Tokyo 113-8656, Japan},
\\ and \name{Minoru \surname{Nohara}}
}

\inst{Department of Physics, Okayama University, Okayama 700-8530, Japan}

\kword{IrTe$_2$, structural phase transition, Rh doping}

\begin{document}
\maketitle

In recent times, IrTe$_2$ has attracted considerable interest because of its peculiar structural/electronic phase transition and the emergence of superconductivity upon chemical doping\cite{Pyon,Yang} or intercalation.\cite{Yang,Kamitani}
IrTe$_2$ crystallizes in a trigonal CdI$_2$-type structure with the space group $P\bar{3}m1$ (No. 164).
Edge-sharing IrTe$_6$ octahedra form two-dimensional IrTe$_2$ layers that are stacked along the $c$-axis. In each layer, Ir atoms are connected to form a regular triangular lattice.
This compound undergoes a first-order structural phase transition at approximately 250 K.\cite{Matsumoto}
Matsumoto \textit{et al.} proposed the average structure below 250 K to be a monoclinic one with the space group $C2/m$ (No. 12), in which the Ir--Ir bond length along the $b$-axis is uniformly reduced\cite{Matsumoto} so that the regular triangular lattice is deformed into an isosceles triangular one.
Recently, Yang \textit{et al.}\cite{Yang} revealed through transmission electron microscopy (TEM) measurements that the low-temperature structure is modulated with a wave vector of ${\bf q} = (1/5, 0, -1/5)$ .
Numerous studies have discussed the importance of the orbital degrees of freedom in Ir $5d$. Pyon \textit{et al.}\cite{Pyon} proposed $t_{2g}$ orbital ordering in a manner analogous to NaTiO$_2$.\cite{Pen}
Yang \textit{et al.}\cite{Yang} assigned a modulated structure to the charge-orbital density wave caused by the orbital-driven Peierls instability in terms of the local density approximation (LDA) calculation.
Ootsuki\cite{Ootsuki} conducted an X-ray photoemission study of the Ir 4$f$ core level and suggested modulation of the charge density at the Ir site in a manner consistent with the orbital density wave.
Other studies have focused on the importance of Te 5$p$ orbitals.
Fang \textit{et al.}\cite{Fang} conducted an LDA calculation for the average structure and suggested the occurrence of band splitting of Te 5$p$ that reduces the kinetic energy.
Oh \textit{et al.}\cite{Oh} proposed that the polymeric Te--Te network in the trigonal phase is destabilized in the modulated low-temperature phase.
Thus, the origin of the structural/electronic phase transition at $\sim$250 K remains unclear.
Interestingly, superconductivity emerges at 3.1 K when the structural/electronic phase transition is suppressed by doping electrons of IrTe$_2$, as demonstrated in Ir$_{1-x}$Pt$_x$Te$_2$,\cite{Pyon} Ir$_{1-x}$Pd$_x$Te$_2$,\cite{Yang} Pd$_x$IrTe$_2$,\cite{Yang} and Cu$_x$IrTe$_2$.\cite{Kamitani}

In this paper, we report that the isovalent Rh doping of IrTe$_2$ suppresses the structural/electronic phase transition and induces superconductivity at 2.6 K.
The doping level of Rh that is necessary for suppressing the transition is three times higher than those of Pt\cite{Pyon} and Pd\cite{Yang}.
Ir$_{1-x}$Rh$_x$Te$_2$ might provide us with a unique opportunity to study the origin of the structural/electronic phase transition in IrTe$_2$.

Polycrystalline samples of Ir$_{1-x}$Rh$_x$Te$_2$ with nominal Rh contents of 0.00 $\le$ $x$ $\le$ 0.30 were synthesized using a solid-state reaction.\cite{Pyon}
Powder X-ray diffraction studies confirmed that the obtained samples had a single phase.
The attempt to synthesize $x=$ 0.50 failed; the pyrite RhTe$_2$ was formed as an impurity phase, indicating that the solubility limit of Rh is between $x=$ 0.30 and 0.50.
Lattice parameters were estimated by the Rietveld refinement using RIETAN-FP program.\cite{Izumi}
The magnetization $M$ was measured using Quantum Design MPMS and SQUID-VSM.

\begin{figure}[t]
\begin{center}
\includegraphics[width=5.25cm]{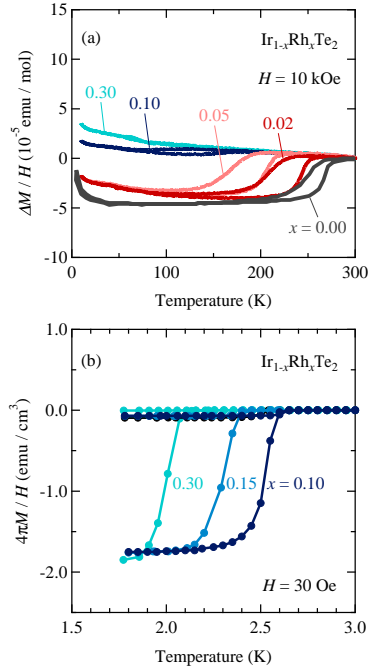}
\caption{
(Color online) (a) Temperature dependence of magnetization $M$ divided by magnetic field $H$ for Ir$_{1-x}$Rh$_x$Te$_2$ at $H=$ 10 kOe upon cooling and heating. The value at 300 K is subtracted for clarity ($\Delta M = M - M_{\rm 300 K}$). (b) Temperature dependence of $M / H$ for Ir$_{1-x}$Rh$_x$Te$_2$ at $H=$ 30 Oe in the zero-field and field cooling conditions. No correction for the diamagnetizing field has been made.
}
\end{center}
\end{figure}
\begin{figure}[t]
\begin{center}
\includegraphics[width=5.25cm]{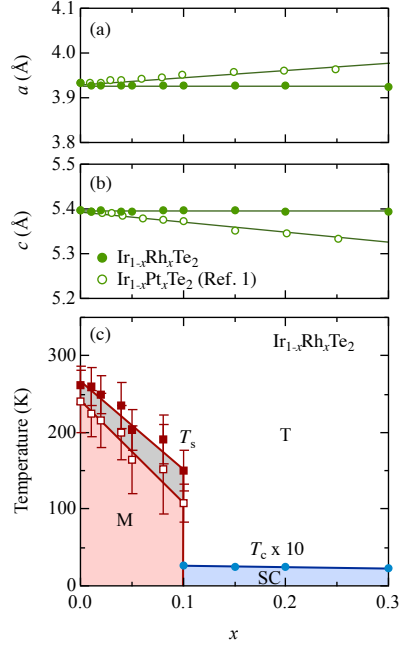}
\caption{
(Color online) (a)(b) Lattice parameters in the trigonal lattice at room temperature as a function of $x$ for Ir$_{1-x}$Rh$_x$Te$_2$. For comparison, those for Ir$_{1-x}$Pt$_x$Te$_2$\cite{Pyon} are also plotted.
(c) Electronic phase diagram of Ir$_{1-x}$Rh$_x$Te$_2$ versus Rh content $x$. T, M, and SC denote the trigonal phase, monoclinic phase, and superconducting phase, respectively. Closed circles show the superconducting transition temperature $T_{\rm c}$ determined from magnetization measurements. For clarity, the values of $T_{\rm c}$ have been scaled by a factor of ten. Closed and open squares show the structural phase transition temperature $T_{\rm s}$ determined from magnetization measurements upon heating and cooling, respectively.
}
\end{center}
\end{figure}
%
Figure 1(a) shows the temperature dependence of magnetization $M$ for Ir$_{1-x}$Rh$_x$Te$_2$ at a magnetic field $H$ of 10 kOe.
$M(T)$ exhibits a hysteretic jump at the first-order structural/electronic phase transition from which we determined the transition temperature
$T_{\rm s}$, as reported previously.\cite{Pyon,Yang,Kamitani,Matsumoto}
$T_{\rm s}$ decreases with increasing $x$ and disappears at $x>$ 0.10.
Superconductivity emerges simultaneously. Figure 1(b) shows the temperature dependence of $M$ at $H=$ 30 Oe.
The considerable shielding signals evidence the emergence of bulk superconductivity in Ir$_{1-x}$Rh$_x$Te$_2$ for 0.10 $\le$ $x$ $\le$ 0.30.
The sample with $x=$ 0.10 exhibits both superconductivity and structural/electronic phase transition. This is likely due to the phase separation at the critical boundary of the transition in the $T=$ 0 limit.
The superconducting transition temperature $T_{\rm c}$ exhibits a maximum of 2.6 K at $x=$ 0.10 and then decreases with increasing Rh content.

The structural/electronic phase diagram of the present system, shown in Fig. 2(c), captures a generic feature of doped IrTe$_2$\cite{Pyon,Yang,Kamitani}: the structural/electronic phase transition is suppressed by chemical doping; the superconducting phase emerges as soon as the transition is lifted; and the superconducting $T_{\rm c}$ exhibits a maximum at the critical boundary of the transition.
However, a distinct difference is observed between the present system and the previous systems with regard to the doping level: a doping of $x$ $\simeq$ 0.1 is necessary to suppress the structural/electronic phase transition for the present Ir$_{1-x}$Rh$_x$Te$_2$ system, whereas a small doping of  $x$ $\simeq$ 0.03 is sufficient for the previous Ir$_{1-x}$Pt$_x$Te$_2$\cite{Pyon} and Ir$_{1-x}$Pd$_x$Te$_2$.\cite{Yang}
The difference could partly be understood by the volume effect:
as shown in Figs. 2(a) and 2(b), the lattice parameters $a$ and $c$ of Ir$_{1-x}$Rh$_x$Te$_2$ are almost independent of doping $x$ and in turn, so is the unit-cell volume.
In Ir$_{1-x}$Pt$_x$Te$_2$, on the other hand, $a$ increases and $c$ decreases with doping $x$, resulting in an increase in the unit cell volume.\cite{Pyon} A similar increase in volume with doping occurs in Ir$_{1-x}$Pd$_x$Te$_2$. \cite{Yang}
Thus, the increase in volume leads to rapid suppression of $T_{\rm s}$.
This is consistent with the observations that $T_{\rm s}$ increases with Se doping\cite{Oh} or the application of hydrostatic pressure\cite{Oh,Kiswandhi}, both of which result in a decrease in the unit cell volume.
In addition, the effect of carrier doping should be considered:
the number of electrons, and in turn the band filling, increases with Pt/Pd doping. This could result in the suppression of a Peierls-type instability as demonstrated in electron-doped Cu$_x$TiSe$_2$.\cite{Morosan}
On the other hand, band filling is unchanged by isovalent Rh doping as long as a rigid-band picture is held.
Thus, both the volume and the doping effects suggest the inability of isovalent Rh doping in suppressing the structural/electronic phase transition of IrTe$_2$.

Nonetheless, our results demonstrate that the structural/electronic phase transition of IrTe$_2$ is suppressed by isovalent Rh doping.
We believe that this observation provides an insight into understanding the structural/electronic phase transition mechanism of IrTe$_2$ at $\sim$250 K.
A clue might be found in the mixed-valent thiospinels CuRh$_2$S$_4$ and CuIr$_2$S$_4$.
CuIr$_2$S$_4$ exhibits a charge disproportionation/ordering transition of Ir at $\sim$230 K\cite{Radaelll} that is considered a charge-orbital density wave.\cite{Khomskii} This transition can be suppressed by isovalent Rh doping.\cite{NMatsumoto}
On the other hand, CuRh$_2$S$_4$ remains a simple metal and exhibits superconductivity at 4.7 K.\cite{Ito}
The same physics may be active in Ir$_{1-x}$Rh$_x$Te$_2$.

In conclusion, the isovalent Rh doping of IrTe$_2$ suppresses the structural/electronic phase transition, resulting in the emergence of superconductivity at 2.6 K.
The doping level of Rh that is necessary to suppress the transition is three times higher than that of other dopants.
Further study of Ir$_{1-x}$Rh$_x$Te$_2$ should provide additional insights that  will help in elucidating the origin of the structural/electronic phase transition in IrTe$_2$.

\begin{acknowledgment}
A part of this work was performed at the Advanced Science Research Center, Okayama University. It was partially supported by Grants-in-Aid for Young Scientists (B) (24740238) and Scientific Research (C) (25400372) from the Japan Society for the Promotion of Science (JSPS) and the Funding Program for World-Leading Innovation R\&D on Science and Technology (FIRST Program) from JSPS.
\end{acknowledgment}

\end{document}